\documentclass[a4paper, 10pt]{article}
\usepackage[T1]{fontenc}
\usepackage[utf8]{inputenc}
\usepackage[english]{babel}

\usepackage{amsmath,amssymb}

\usepackage{graphicx}
\usepackage{bm}
\usepackage[dvipsnames]{xcolor}
\usepackage{siunitx}
\usepackage{dsfont}

\usepackage{authblk}
\usepackage{hyperref}
\usepackage{orcidlink}
\usepackage{csquotes}

\usepackage[sorting=none,citestyle=numeric-comp, natbib]{biblatex}
\bibliography{biblio}

\makeatletter
\newlength{\negph@wd}
\DeclareRobustCommand{\negphantom}[1]{%
  \ifmmode
    \mathpalette\negph@math{#1}%
  \else
    \negph@do{#1}%
  \fi
}
\newcommand{\negph@math}[2]{\negph@do{$\m@th#1#2$}}
\newcommand{\negph@do}[1]{%
  \settowidth{\negph@wd}{#1}%
  \hspace*{-\negph@wd}%
}
\makeatother

\newcommand{\Prob}{\mathbb{P}}
\newcommand{\AUL}{A_{\text{\textsc{ul}}}}
\newcommand{\TUL}{\vartheta_{\text{\textsc{ul}}}}

\newcommand{\msub}{_{[m]}}

\begin{document}

\title{Near-exact Upper Confidence Limits\\ for Axion Haloscopes}

\author[1]{Andrea Gallo Rosso~\orcidlink{0000-0002-4664-5504}%
\thanks{\href{mailto:andrea.gallo.rosso@fysik.su.se}{andrea.gallo.rosso@fysik.su.se}}}

\author[1]{Jan Conrad~\orcidlink{0000-0001-9984-4411}%
\thanks{\href{mailto:conrad@fysik.su.se}{conrad@fysik.su.se}}}

\affil[1]{\small Stockholm University and The Oskar Klein Centre
for Cosmoparticle Physics,\\ Alba Nova, Stockholm, Sweden}

\date{}

\maketitle

\begin{abstract}
The data acquisition of axion haloscopes commonly implements sequential adaptive frequency scans.  We present a method to calculate likelihood-based upper confidence limits on axion-photon coupling under this circumstance. We compare the likelihood-based method conceptually and quantitatively with tailor-made methods, i.e.\ the one  proposed by Brubaker \textit{et al} (Geometric) and its upgrade the Bayesian Power-Measured (BPM) framework proposed by Palken \textit{et al}. We find that the likelihood method yields average upper limits that are significantly more constraining than those obtained from these methods. For example, expected constraints improve by a factor of 1.4 as compared with the current state of the art when applied to projections for the ALPHA Plasma-haloscope. We further clarify the statistical basis for the Geometric and BPM methods. In particular, we point out the correspondence of BPM to e-values, proposed as an alternative to conventional p-values in statistics.
\end{abstract}

\section{Introduction}

The nature of dark matter is one of the most pressing problems in modern fundamental physics \cite{Bergstrom:2000pn,Bertone:2004pz,Jaeckel:2010ni,Feng:2010gw,Profumo:2019ujg,Arbey:2021gdg,Green:2022hhj}. Among the many particle candidates, axions are of particular interest. Their existence is motivated by symmetry considerations in the strong interactions \cite{Peccei:1977hh, Peccei:1977ur, Weinberg:1977ma, Wilczek:1977pj, Kim:1979if, Shifman:1979if, Dine:1981rt, Zhitnitsky:1980tq}, and well motivated production mechanisms imply that axions could constitute dark matter\cite{Preskill:1982cy,Abbott:1982af,Dine:1982ah}.
Possible axion masses span 19 orders of magnitude, so an abundance of experimental techniques is applied to find them \cite{Irastorza:2018dyq, Billard:2021uyg, Irastorza:2021tdu, Semertzidis:2021rxs, Adams:2855525, AxioBook2022,Berlin:2024pzi}.

A particularly promising venue is provided by  ``haloscopes'' \cite{Sikivie:1983ip,Sikivie:1985yu}---cavities that can detect the power deposited by an axion conversion to photons prompted by a strong magnetic field. Haloscopes operate by scanning a frequency range for a narrow-band excess at an unknown frequency $\nu_a$, corresponding to the axion mass $m_a\approx h \nu_a/c^2$.
Haloscope analysis protocols \citep[e.g.,][]{Brubaker:2017ohw,Brubaker:2017rna,Foster:2017hbq,Knirck:2018knd,Palken:2020wgs,Ahn:2020nyv,GalloRosso:2022mhx} have the possibility of acquiring additional data at promising frequencies. A power excess that meets a certain threshold condition triggers one (or more) additional measurements, also called ``rescans.''
This is a very useful feature for axion searches, but poses also a challenge to statistical inference, as data acquisition is adaptive and conditioned on the observed data itself.

\section{Upper Limits in axion haloscopes}
In haloscopes, the observed excess $X_{\nu}$ at a frequency $\nu$ is Gaussian with very good approximation \cite{Brubaker:2017rna,Palken:2020wgs}. Its mean is null if no axion is present; otherwise, it is proportional to the axion-photon coupling, henceforth generally denoted as $\mathbb{A}$ (or $A$) for the true (or tested) value:
\begin{equation}
    X_\nu\sim\mathcal{N}(\mathbb{A}u_\nu, \sqrt{u_\nu}).
\end{equation}
Here, $u_{\nu} > 0$ is an alias for the signal-to-noise ratio at $\nu$, given the specific experimental setting and measurement time (see also \cite{GalloRosso:2022mhx}). With this notation, the ($i+1$)\textsuperscript{th} rescan is triggered whenever $X_{\nu, i}$ satisfies $X_{\nu,i}\ge b_{\nu,i}$, where $b_{\nu,i}$ is a threshold commonly defined in terms of the system noise. After assessing the presence of an axion signal (rejection of noise-only hypothesis) anywhere in the frequency range, statistical inference turns to exclusion. It consists of the point-wise statement on the maximal coupling $\mathbb{A}$ at a given frequency $\nu$ still consistent with the observation at a certain confidence level (CL).
Since exclusion is defined pointwise in frequency, we will omit the subscript $\nu$ from now on.

In the former case the null hypothesis is noise, in the latter it is $\mathbb{A} > \AUL$. Conflating the two can lead to unwanted results. For example, the protocol implemented by Brubaker \textit{et al.} and used in early HAYSTAC analysis \cite{Brubaker:2017ohw,Brubaker:2017rna} applies a geometric test with binary outcome (signal/no-signal) defined by the local $5\sigma$ significance for the false-negative rate of a given discovery benchmark $A_{\text{\textsc{t}}}$. The latter corresponds to the prescribed CL at each stage thanks to a calibration of the $u_i$ values.
The same $A_{\text{\textsc{t}}}$ becomes the quoted value for exclusion when no evidence is found, and the same nominal exclusion CL is assigned to candidates failing any of the scans.

However, this argument conditions on passing the previous scan(s).
A frequentist exclusion must be calibrated over the full decision tree (i.e., all possible experimental outcomes) whose total false negative probability is $F_n = f_1+(1-f_1)f_2$, with $f_i = \Phi[(b_i - Au_i)/\sqrt{u_i}]$, where $\Phi$ is the standard normal CDF.

In general, the geometric upper limit $\AUL^{g_{N_s}}$ satisfies:
\begin{equation}\label{eq:def:geomUL}
    \alpha = \sum_{i=1}^{N_s}\frac{f_i(\AUL^{g_{N_s}})}{1-f_i(\AUL^{g_{N_s}})}
    \prod_{k=1}^{i}\left[1-f_k(\AUL^{g_{N_s}})\right],
\end{equation}
where $N_s$ is the maximum number of allowed scans.
Notably, definition \eqref{eq:def:geomUL} implies that the upper limit weakens as $N_s$ increases, since it raises the cumulative false-negative probability. Moreover, while its binary nature protects against downward fluctuations, it ignores information from previous scans. Hence, the result becomes overly conservative and closer to an \emph{a priori} sensitivity benchmark.

In addition, exclusion \eqref{eq:def:geomUL} and frequentist confidence intervals in general require that the protocol be fully specified in advance. This does not include just the maximum number of allowed rescans $N_s$, but also the thresholds $b_i$ and the conditions $u_i$ during rescans that were never performed. Incidentally, this is also true for the likelihood method that we are proposing here, though as we will see the effect of conservative assumptions on the ensemble of experiments is minor.

The Bayesian Power-Measured (BPM) framework proposed by Palken \textit{et al.}\ \cite{Palken:2020wgs} tackles this issue by means of a hybrid approach between Bayesian and frequentist inference.  
For each scan $i$ performed and power excess $X_i$ collected, equation (9) of Ref.\ \cite{Palken:2020wgs} defines:
\begin{equation}\label{eq:def:Lambda}
	\Lambda_i(A)\equiv \frac{\Prob(X_i\mid A)}{\Prob(X_i\mid A=0)}.
\end{equation}
For sequential independent measurements, evidence collected through the scans accumulates multiplicatively through the total update $\Lambda(A)=\prod_i\Lambda_i(A)$---this, regardless of any specifications on $N_s$, $b_i$, or $u_i$.

This quantity can be interpreted both in frequentist and Bayesian context; in particular, it can be viewed as both a likelihood ratio \cite{NeymanPearson1933} and as a Bayes factor of $A$ against the null in the limit of infinitesimal priors. It can also provide evidence either for or against a specified signal strength in both Bayesian \cite[e.g.][]{jeffreys1998,kass1995} and frequentist frameworks. In the latter, test and rejection of hypotheses on $A$ are based on a threshold condition on $\Lambda(A)$. Ref.\ \cite{Palken:2020wgs} connects the Bayesian interpretation to frequentist exclusion through the geometric (BT2) framework. The result is then extended to BPM, but without directly calibrating its false-exclusion probability.

In fact, in the BT2 framework $\Lambda^{\scriptscriptstyle \mathrm{BT2}}$ can take only two values (success/fail). As equation (A8) of Ref.\ \cite{Palken:2020wgs} shows, in a small false-positive regime $\Lambda_{\rm fail}^{\scriptscriptstyle \mathrm{BT2}}$ approximates the total false negative rate $F_{\rm n}$. Hence, $\Lambda_{\rm fail}^{\scriptscriptstyle \mathrm{BT2}}=0.1$ approximately corresponds to $90\%$ exclusion. In BPM,
$\Lambda$ becomes a ratio of continuous likelihoods. While it is argued that the interpretation of the update is unchanged, $\Lambda\simeq F_{\rm n}$ relies on the binary space and does not necessarily extend to continuous power measurements. The normalization of $\Lambda$ provides a shared scale for the relative evidence across BT2 and BPM, but not for the error probability.

This becomes evident by realizing the connection of BPM and the formalism of e-variables \cite[e.g.][]{Grunwald:2024,RamdasWang:2025}, justified by the fact that $\mathbb{E}_0[\Lambda_i(A)]=1$. The realization of an e-variable (i.e., e-value) is safe against type-I and -II errors thanks to the Markov inequalities:
\begin{equation}\label{eq:Mkv:0}
    \Prob_0\left(\Lambda_i(A)\ge1/\alpha\right)\le\alpha
    \quad\text{and}\quad
    \Prob_A\left(\Lambda_i(A)\le\lambda\right)\le\lambda,
\end{equation}
with $\alpha$, $\lambda\in(0,1]$. Relations \eqref{eq:Mkv:0} apply also to the ensemble $\Lambda(A)$, provided each $\Lambda_i(A)$ shares the same null, is computed on uncorrelated data, and satisfies $\mathbb{E}_0[\Lambda(A)]\le 1$ conditioned on the result of the previous scan \cite{Ville:1939, Grunwald:2024}. 

As a consequence, $\Lambda(A)\le\lambda$ guarantees a false-exclusion probability no larger than $\lambda$, but the bound is not necessarily saturated. For instance, a 1-scan measurement has a maximal false-exclusion probability $\Phi(-\kappa)$ with $\kappa = \sqrt{2\ln(1/\lambda)}$. This  probability is only $1.6\%$ for $\lambda=0.1$, rather than 10\%.

\begin{figure}[t]
\centering
\includegraphics[width=0.75\textwidth]{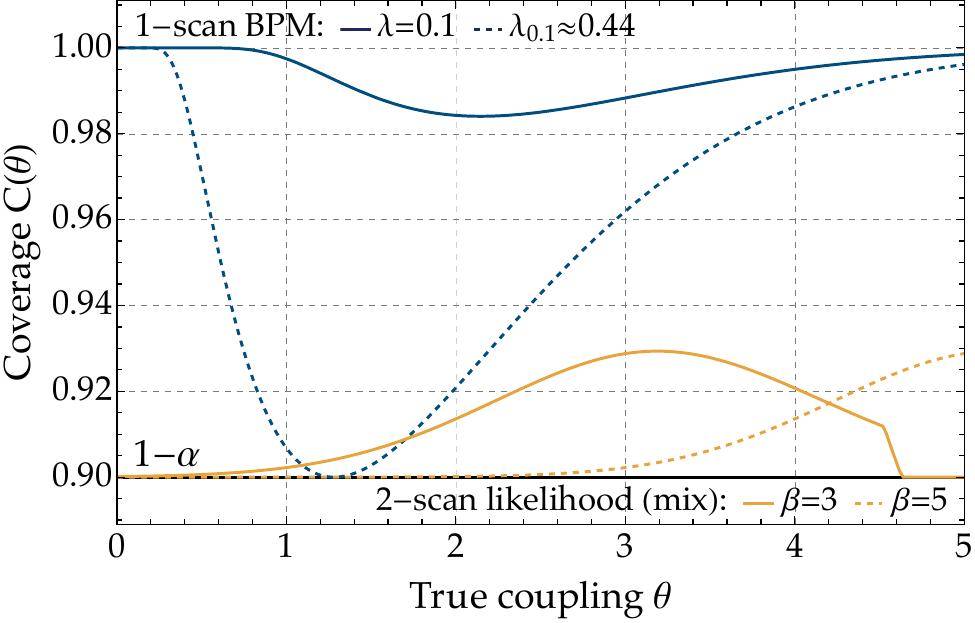}
\caption{Coverage as a function of $\theta=\mathbb{{A}}\sqrt{u_1}$ for the 1-scan BPM and the 2-scan likelihood ``mix'' frameworks.
Calibrated value $\lambda_{0.1}$ for BPM comes from equation \eqref{eq:calk}---see text for details.
\label{fig:cov}
}
\end{figure}

In general, $\lambda_\alpha$ (or $\kappa_\alpha$) can be calibrated in a frequentist sense, such that
\begin{equation}\label{eq:calk}
     \lambda_\alpha = \sup\left\{\lambda>0: \sup_{\mathbb{A}>0}\Prob_\mathbb{A}(\Lambda(\mathbb{A})\le\lambda)\le\alpha \right\}.
\end{equation}
This prescribes a minimal coverage of $(1-\alpha)$ at the least-favorable $\mathbb{A}$ at the expenses of procedural freedom.

\section{Likelihoood-based upper limits}
\label{sec:hypo}

If procedural freedom is to be sacrificed, a frequentist construction can provide an exact exclusion probability. The starting point of the framework presented here is the same as BPM---that is, a series of (re)scanned power excesses $X_i$ provided e.g.\ by the application of a likelihood-ratio test \cite[e.g.][]{Foster:2017hbq,GalloRosso:2022mhx} or sequential transformations on the power spectra \cite[e.g.][]{Brubaker:2017ohw,Brubaker:2017rna}.

Following Ref.\ \cite{GalloRosso:2022mhx}, we define cumulative variables:
\begin{equation}\label{eq:def:S}
    S_m = \frac{\sum_{i=1}^{m} X_i}{\sqrt{\sum_{i=1}^{m} u_i}}=
    \frac{X\msub}{\sqrt{u\msub}}\sim\mathcal{N}(\theta_m,1),
\end{equation}
so that the trigger condition $S_m\ge\beta_m$ automatically includes prior evidence from the $(m-1)$ previous scans:
\begin{equation}\label{eq:trig:Si}
    S_m\ge\beta_m \quad\Longleftrightarrow\quad 
    X\msub\ge \beta_m \sqrt{u\msub}.
\end{equation}
The variables $S_m$ are correlated according to
\begin{equation}\label{eq:def:rho}
    \rho_{jk}
    \equiv
    {\rm Cov}(S_j,S_k) =
    \sqrt{\frac{u_{[\min(j,k)]}}{u_{[\max(j,k)]}}}.
\end{equation}

Couplings $\theta_m=\mathbb{A}\sqrt{u\msub}$ and $\vartheta_m=A\sqrt{u\msub}$ are not independent parameters, but branch-dependent representations of $\mathbb{A}$ and $A$. Hence, we define the dimensionless strengths $\theta\equiv\theta_1$ and $\vartheta\equiv\vartheta_1$, so that $\theta_m = \theta\sqrt{u\msub/u_1}$.

To test a value $\vartheta$ given $M=m$ performed scans, an observed $S_m=s_m$, and an unconstrained maximum-likelihood estimator (MLE) $\hat\vartheta_m=s_m$, we use the one-sided likelihood-ratio statistic:
\begin{equation}\label{eq:def:qa}
    q_\vartheta = q_\vartheta(\hat{\vartheta}_m, m) =
    \begin{cases}
        (\vartheta_m-\hat\vartheta_m)^2, & \hat\vartheta_m\leq\vartheta_m;\\
        0, & \hat\vartheta_m > \vartheta_m.
    \end{cases}
\end{equation}
The case in which $\vartheta$ is constrained to be non-negative is described by the statistic $q_{\vartheta}^{+}$ defined in \eqref{eq:def:qa:const}.

The $p$-value is given by the sum over all the $N_s$ mutually exclusive stopping stages $m$:
\begin{equation}\label{eq:def:pVal}
    p_\vartheta = \sum_{m=1}^{N_s} \Prob_\vartheta\left(M = m \,, q_\vartheta(\hat{\vartheta}, m) \geq q_{\vartheta}^{\mathrm{obs}}\right),
\end{equation}
where $\Prob_{\vartheta}$ denotes probabilities evaluated assuming the tested signal strength $\vartheta$, and $q_{\vartheta}^{\rm obs}$ is the statistic evaluated on the observed experimental outcome.

Since $q_{\vartheta}$ (and $q_{\vartheta}^+$) are monotonically decreasing for $\hat\vartheta_m\leq\vartheta_m$, condition \eqref{eq:def:pVal} translates as a single threshold on $\hat\vartheta_m$, whose expression is given in Appendix \ref{app:pval}.

For $N_s=1$, equation \eqref{eq:def:pVal} reduces to
\begin{equation}
    p_\vartheta^{(1)}
    =
    P_\vartheta(S_1\leq s_1)
    =
    \Phi(s_1-\vartheta).
    \label{eq:p1}
\end{equation}
Imposing $p_\vartheta^{(1)}\equiv\alpha$ gives the upper limit at $(1-\alpha)$ CL: $\TUL^{\ell1} = s_1+\Phi^{-1}(1-\alpha)$ for the unconstrained case, and $\TUL^{\ell 1+}=\max\!\left(0,\vartheta_{\rm UL}^{\ell 1}\right)$ for the constrained one.

For $N_s=2$, it is convenient to define $B_m(\vartheta)\equiv\beta_m-\vartheta_m$ and $R_m(\vartheta)\equiv r_m(\vartheta)-\vartheta_m$, with $r_m(\vartheta)$ defined in \eqref{eq:def:rm}.
The $p$-value is then given by the close form:
\begin{equation}
    p_\vartheta^{(2)}=\Phi\!\left[\min(B_1,R_1)\right]+
    \Phi(R_2)-\Phi_2(B_1,R_2;\rho_{12}),
    \label{eq:def:p2}
\end{equation}
where $\Phi_2$ is the standard bivariate normal CDF with correlation $\rho_{12}$ \eqref{eq:def:rho}.
Derivation of equation \eqref{eq:def:p2} and its extension to $N_s>2$ are given in the Appendix.

The frequentist two-scan upper limit is defined implicitly by imposing equation \eqref{eq:def:p2} to be equal to $\alpha = 1 - \rm CL$. The same relation applies to the constrained case, provided that $R_m^+(\vartheta)\equiv r_m^+(\vartheta)-\vartheta_m$ is used. 

To characterize the behavior of this framework, we consider the expected upper limit, $\mathbb{E}[\TUL]$, which is computed in Appendix \ref{app:expected}.
We also define the dimensionless quantities $\beta = b_1/\sqrt{u_1}$ and $\omega=u_2/u_1$. The former defines the rescan trigger; the latter controls the amount of information from a possible second scan: $\omega\to0$ recovers the one-scan result (uninformative second scan) while $\omega\to\infty$ represent a second scan of arbitrary precision.

\begin{figure}[t]
\centering
\includegraphics[width=0.8\textwidth]{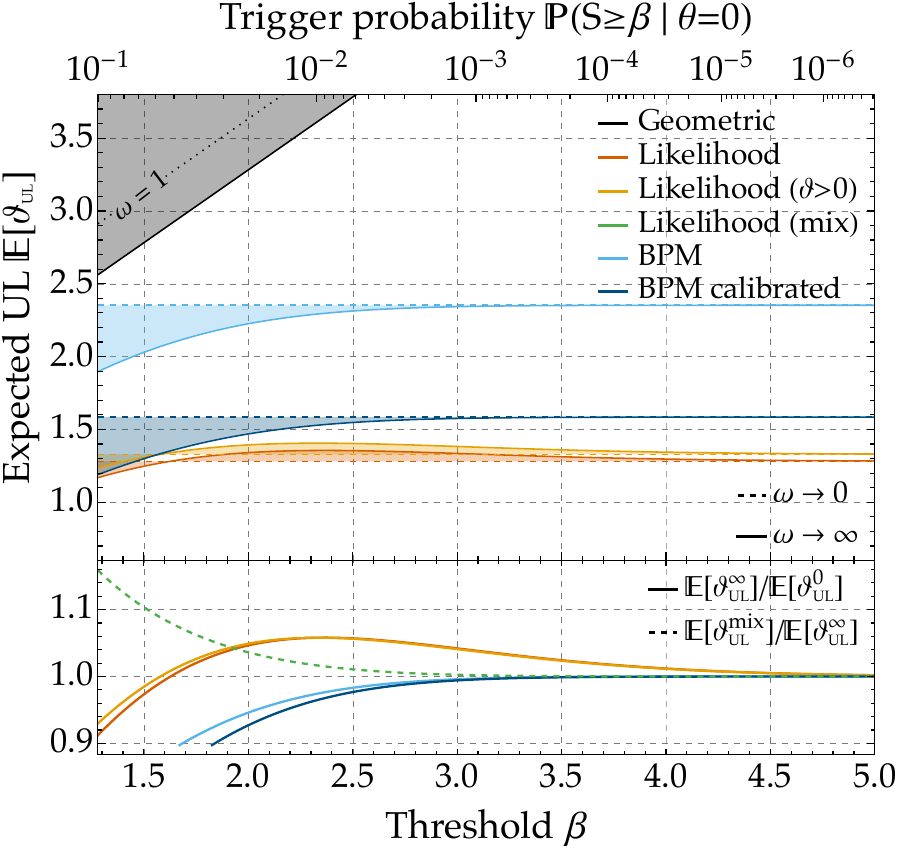}
\caption{Top: Expected upper limit as a function of $\beta$, for $\theta = 0$ and $90\%$ CL. Bottom: Relative difference between the limiting cases $\omega\to0$ and $\omega\to\infty$ (solid) and between the ``mix'' and $\omega\to\infty$ unconstrained likelihood approach (dashed). The geometric test reaches $\mathbb{E}[\TUL] = 6.3$ for $\beta = 5$.
\label{fig:bUL}
}
\end{figure}

For $N_s = 2$ and $\omega\to\infty$, experiments that stop after the first scan incur a penalty from the unrealized branch with respect to the one-scan-only case, while realizations that trigger a rescan benefit from its arbitrary precision. Therefore, $\mathbb{E}[\TUL^{\ell2}]<\mathbb{E}[\TUL^{\ell1}]$ if $\beta$ is low enough to favor rescans. If not, $\mathbb{E}[\TUL^{\ell2}]>\mathbb{E}[\TUL^{\ell1}]$ and $\mathbb{E}[\TUL^{\ell2}]\to\mathbb{E}[\TUL^{\ell1}]$ as $\beta$ increases, since rescans become increasingly rare and the probability assigned to the second branch vanishes. In general, in the regime of rare rescans, $\mathbb{E}[\TUL]$ weakens for increasing $N_s$, similarly to the geometric construction.

Since haloscopes operate in this latter regime, we can define a practical recipe (labeled ``mix'') to deal with the unknown unrealized $u_i$: if $m<N_s$ scans are observed, quote the upper limit for $N_s=m+1$  and $u_{m+1}\to\infty$.
Under the assumptions of Figure \ref{fig:bUL}, the relative difference induced by $\omega$ in the likelihood upper limits is no higher than $6\%$ for $\beta\gtrsim2$, with the ``mix'' recipe worsening the un/constrained frameworks by a further 3.6\% (0.4\%) for $\beta\ge2$ ($\beta\ge3$) at most.

The ``mix'' framework is conservative also in its coverage (see Appendix \ref{app:coverage}). As Figure \ref{fig:cov} shows for $N_s = 2$, $C_{\rm mix}^{(2)}(\theta)$ deviates from the nominal CL until both 1-scan rejections thresholds $R_1$ exceed the stopping threshold $B_1$---see equation \eqref{eq:CovN2}. Moreover, under constraint $\theta\ge 0$, coverage is unity at the boundary $\theta = 0$. The transition is continuous, with changes in slope caused by these threshold crossings. For more than two scans, the coverage is determined by the sum of the branch contributions in equation \eqref{eq:Cmix:Ns}, and the conservatism must be assessed for the chosen protocol.

\section{Comparison among frameworks}

In this section we extend the comparison to the other frameworks, while keeping $N_s\le 2$. Similar considerations hold for greater $N_s$.

In the geometric framework, the one-scan limit corresponds to $\omega\to\infty$, when a nonzero signal would trigger a positive experimental outcome (success) with probability 1. 
In order not to introduce an additional second threshold $b_2$ just for this framework, we assign the same false-positive rate $\beta$ to both scans. This choice matches the spirit of Refs.\ \cite{Brubaker:2017ohw,Brubaker:2017rna} and experimental practice, but it causes $\omega\to 0$ to be singular. A finite threshold cannot be overcome by a second scan carrying no information. However, this regime is of little experimental relevance.

For BPM, some care is needed in defining an equivalent two-scan comparison, since its e-value construction does not require the stopping rule to be fixed in advance.  We therefore consider two implementations. The first one, labeled ``fix,'' follows the same stopping protocol as the likelihood-based construction, and the condition $\Lambda(\vartheta)\leq\lambda$ is evaluated only at the end, when the procedure halts. In the second one, labeled ``any,'' the same condition is checked after every acquisition, and a value is excluded as soon as the cut is crossed.

The expected upper limit $\mathbb{E}[\TUL]$ for the different frameworks, as a function of $\beta$, for $\theta = 0$ and $90\%$ CL is shown in Figure \ref{fig:bUL}. The likelihood framework considers both the constrained and unconstrained case. The BPM upper limit, on the other hand, has been computed both for $\lambda = 1-\rm CL$ and with coverage calibrated according to \eqref{eq:calk}. In the specified regime, $\vartheta\le\vartheta_c\equiv\beta+\sqrt{\beta^2+\kappa^2}$ (see Appendix \ref{app:expected}) therefore BPM-fix = BPM-any $\equiv$ BPM. The BPM framework gives an upper limit which is $\sim 20\%$ higher than the likelihood one when the coverage is calibrated, up to $\sim 80\%$ higher when it is not.

\begin{figure}[t]
\centering
\includegraphics[width=0.8\textwidth]{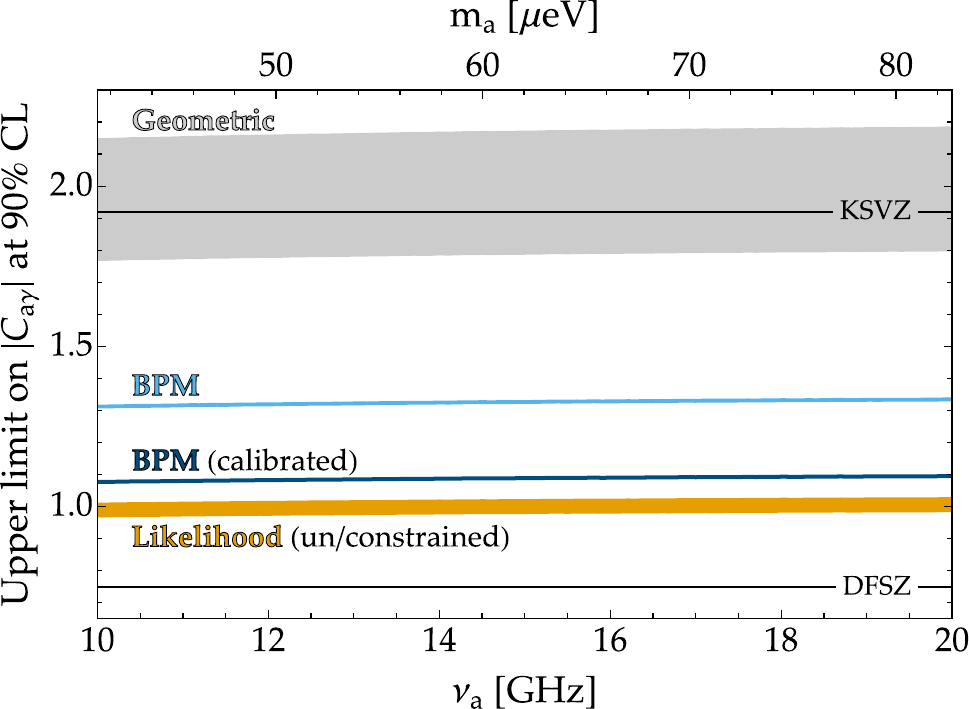}
\caption{Expected upper limits at $90\%$ CL on the $|C_{a\gamma}|$ coupling achievable by ALPHA \cite{ALPHA:2026vwc} (see text). %
The line-widths/shaded regions represent the interval $3\le\beta\le 5$.}
\label{fig:ALPHA}
\end{figure}

In order to provide also a physical benchmark, Fig.~\ref{fig:ALPHA} shows the upper limit achievable by ALPHA (Axion Longitudinal Plasma HAloscope) targeting the 10–20 GHz (40--\SI{80}{\micro\electronvolt}) range while operating at the conditions discussed in Ref.\ \cite{ALPHA:2026vwc}. 
The statistical assumptions are those adopted throughout this section, with $\omega\to\infty$, and $\beta$ from 3 to 5. Here, the uncalibrated BPM (Geometric) upper limits are 30\% (115\%) higher (i.e.\ less constraining) compared to the Likelihood framework, while the application of the ``mix'' recipe would increase their limits by $\sim 0.5\%$.

\section{Conclusions}
Statistical inference for axion haloscopes has one noteworthy feature: the possibility of sequentially acquiring new data in so-called rescans. In this work, we showed that a conventional likelihood approach yields substantially stronger expected limits.

The challenge of frequentist inference for axion haloscopes is specifying in advance all possible outcomes of the experiment, including stopping rules and detector configurations for scans that may never be performed. A fully Bayesian approach, defining the appropriate Bayes factor, is a viable  solution. However, insisting on frequentist coverage may yield overly conservative results.

A fully frequentist approach is also a viable solution.
We showed that, in experimental regimes, uncertainty about future detector configurations has only a limited impact, even when all the possible experimental outcomes are included. We also proposed a practical ``mix'' prescription to reduce dependence on unrealized detector configurations while keeping its statistical performance explicitly assessable.

We performed explicit quantitative comparisons for $N_s\le2$, while the likelihood construction and coverage expressions extend to $N_s>2$ with the same conclusions. In regimes of rare rescans, the contribution by $p$-values from higher scans becomes negligible by construction. Future work will address the problem of discovery. 

\section*{Acknowledgments}
The authors thank Jón Gu\dh mundsson, Junu Jeong, Andrew Meyer, and Michael Sekatchev for their feedback and comments. The authors acknowledge valuable discussions within the ALPHA Collaboration.
The authors acknowledge support from the Knut and Alice Wallenberg Foundation and Olle Engkvists Foundation. AGR acknowledges support from the EDUCATE excellence centre funded by the Swedish Research Council through grant Dnr 2022-06627, and  the use of Claude (Anthropic) and ChatGPT (OpenAI)---as a help to brainstorm and support the central idea and the theoretical framework, to assist in coding the numerical studies, and to edit the manuscript. The authors take full responsibility for the content of this work.

\appendix

\section[Likelihood-Ratio tests and p-values]{Likelihood-Ratio tests and $\bm p$-values}
\label{app:pval}

In this section, we derive the likelihood-ratio statistics, the corresponding thresholds on the scan-dependent maximum-likelihood estimator and the explicit expression for the $p$-values.

For an outcome consisting of $M=m$ scans, with an observed value $S_m = s_m$, as per \eqref{eq:def:S}, the log-likelihood is: $\ell_m(\vartheta)=-(s_m-\vartheta_m)^2/2+{\rm const}.$

The likelihood-ratio statistic $q_\vartheta=2[\ell(\hat\vartheta_m)-\ell(\vartheta_m)]$ has then the form given in \eqref{eq:def:qa}. On the other hand, if the physical parameter is constrained to be non-negative, the MLE becomes $\tilde\vartheta_m= \max(0,\hat\vartheta_m)$, which defines
\begin{equation}\label{eq:def:qa:const}
    q_\vartheta^+ =
    \begin{cases}
        (\vartheta_m^2 - 2\vartheta_m\hat\vartheta_m), & \hat\vartheta_m < 0;\\
        (\vartheta_m - \hat\vartheta_m)^2, & 0 \leq \hat\vartheta_m\leq\vartheta_m;\\
        0, & \hat\vartheta_m > \vartheta_m.
    \end{cases}
\end{equation}

The condition $q_\vartheta\ge q_\vartheta^{\rm obs}$ defining the p-value \eqref{eq:def:pVal} can be expressed as
\begin{equation}\label{eq:def:rm}
    \hat\vartheta_m \leq r(\vartheta_m) \equiv\vartheta_m - \sqrt{q_\vartheta^{\mathrm{obs}}},
\end{equation}
while the constrained case gives
\begin{equation}
    \hat\vartheta_m \leq
    r^+(\vartheta_m)\equiv
    \begin{cases}
    r(\vartheta_m),
    & q_\vartheta^{\rm obs}\le \vartheta_m^2;\\
    \dfrac{\vartheta_m}{2}-\dfrac{q_\vartheta^{\rm obs}}{2\vartheta_m},
    & q_\vartheta^{\rm obs}>\vartheta_m^2 .
    \end{cases}
\end{equation}

The closed form in equation \eqref{eq:def:p2} follows directly by separating the mutually exclusive stopping branches; that is, for $N_s=2$: $p_\vartheta^{(2)}=p_{\vartheta,1}^{(2)}+p_{\vartheta,2}^{(2)}.$ Using the standardized quantities $B_m$, $R_m$, and $Z_m\equiv S_m -\vartheta_m$, we get:
\begin{equation}
    p_{\vartheta,1}^{(2)}
    =
    \Prob_\vartheta
    \left(Z_1<B_1,Z_1\leq R_1
    \right) = 
    \label{eq:p21_def}
    \Phi\!\left[\min(B_1,R_1)\right].
\end{equation}
For the second branch, we have
\begin{align}\label{eq:p2:cpr}
    p_{\vartheta,2}^{(2)}
    &=
    \Prob_\vartheta
    \left(
        Z_1\geq B_1,\,
        Z_2\leq R_2
    \right)\\
    &= \Prob_\vartheta(Z_2\leq R_2)
    -
    \Prob_\vartheta
    (Z_1<B_1,Z_2\leq R_2)\\
    &=
    \Phi(R_2)
    -
    \Phi_2(B_1,R_2;\rho_{12}).
    \label{eq:p22}
\end{align}

The same construction extends recursively to an arbitrary number
of possible scans. For $N_s=3$,
the first contribution remains unchanged: $p_{\vartheta,1}^{(2)}=p_{\vartheta,1}^{(3)}$.
The second branch is reached only after the first threshold has
been exceeded and the second threshold has not,
while at the same time satisfying the likelihood-ratio ordering:
\begin{align}
    p_{\vartheta,2}^{(3)}
    &=\label{eq:p32}
    \Prob_\vartheta
    \left(
        Z_1\geq B_1,\,
        Z_2\leq\min(B_2,R_2)
    \right)\\
    &=
    \Phi\!\left[\min(B_2,R_2)\right]
    -\Phi_2\!\left[B_1,\min(B_2,R_2);
        \rho_{12}\right].
\end{align}

Finally, reaching the third scan requires the first two thresholds
to have been exceeded. This gives:
\begin{align}
    p_{\vartheta,3}^{(3)}
    =&
    \Prob_\vartheta
    \left(
        Z_1\geq B_1,\,
        Z_2\geq B_2,\,
        Z_3\leq R_3\right)\\
    =&\,\Phi(R_3) - \Phi_2(B_1,R_3;\rho_{13})
    -\Phi_2(B_2,R_3;\rho_{23})+
    \Phi_3\!\left(
        B_1,B_2,R_3;
        \bm{\rho}_{123}
    \right),
    \label{eq:p33}
\end{align}
where we applied inclusion--exclusion \cite[e.g.][]{Feller:1968,Genz:2009}; $\Phi_3$ is the trivariate standard-normal CDF and
$\bm{\rho}_{123}$ is the correlation matrix of elements $[\bm{\rho}_{123}]_{ij}=\rho_{ij}$ as per \eqref{eq:def:rho}.

More generally, the contribution from a branch terminating at scan $m$ requires the trigger from all the previous $(m-1)$ thresholds, while preserving the likelihood-ratio condition \eqref{eq:def:pVal}. Therefore, the contribution of that branch is a joint probability of the series $(Z_1,\ldots,Z_m)$, evaluated in terms of Gaussian CDFs, and using the inclusion-exclusion principle to account for the previous $(m-1)$ trigger conditions. The constrained construction ($\theta\ge0$) follows identically after replacing $R_m$ by $R_m^+$.

\section{On the coverage}
\label{app:coverage}

In this section, we apply the branch-related $p$-values to the computation of the coverage of the ``mix'' framework.

Coverage is defined as $C(\theta) = \Prob_\theta\!\left(\TUL^{\rm mix}\ge\theta\right)$. Let $p_{\theta,m}^{(N_s)}(q)=\Prob_\theta(M=m,\,q_\theta>q)$ be the $m^{\rm th}$ branch contributions evaluated at a generic threshold $q$, under the full frequentist protocol at $N_s$ allowed scans.

When the accepted $\vartheta$ are all those $\vartheta\le\TUL^{\rm mix}$, noncoverage at the true value $\theta$ is equivalent to rejecting $\theta$. Within branch $M=m$, rejection occurs for $q_\theta>q^{\rm crit}$, or equivalently $Z_m<R_m(\theta;q_m^{\rm crit})$. For $m=N_s$, $q^{\rm crit}_{m}(\theta)$ coincides with $q^{\rm full}_{m}(\theta)$ from the full protocol: $\sum_i^{N_s} p_{\theta,i}^{(N_s)}(q_{N_s}^{\rm full})=\alpha$. On the other hand, and for $m<N_s$, $q^{\rm crit}(\theta)$ satisfies
\begin{equation}\label{eq:qcrit}
    \lim_{u_{m+1}\to\infty}\sum_{j=1}^{m+1}
    p_{\theta,j}^{(m+1)}
    \!\left(q^{\rm crit}_{m}\right) = \alpha.
\end{equation}

Then, summation over the mutually exclusive stopping branches gives
\begin{equation}
    C_{\rm mix}^{(N_s)}(\theta)=1-\sum_{m=1}^{N_s}
    p_{\theta,m}^{(N_s)}(q^{\rm crit}_{m}),
\end{equation}
where $p_{\theta,m}^{(m+1)}(q^{\rm crit}_{m})\equiv p_{\theta,m}^{(N_s)}(q^{\rm crit}_{m})$ for $m<N_s$, since both protocols have identical stopping conditions and cumulative statistics through scan $m$.

We can exploit again the exact coverage of the full construction, which gives
\begin{equation}\label{eq:Cmix:Ns}
    C_{\rm mix}^{(N_s)}(\theta)=1-\alpha+ \!\sum_{m=1}^{N_s-1}\!
    \left[ p_{\theta,m}^{(N_s)}(q_{N_s}^{\rm full}) -p_{\theta,m}^{(N_s)}(q^{\rm crit}_{m})
\right].
\end{equation}
The same reasoning apply to the constrained construction ($\theta\ge 0$). At the physical boundary, $C_{\rm mix}^{(N_s)}(0)=1$. If the full calibration is conservative rather than exact, $C_{\rm full}^{(N_s)}(\theta)$ replaces $1-\alpha$.

For $N_s = 2$, relation \eqref{eq:Cmix:Ns} becomes
\begin{equation}\label{eq:CovN2}
    \begin{aligned}
        C_{\rm mix}^{(2)}(\theta) =& 
        1-\alpha+\Phi\!\left[\min\!\left(B_1(\theta),R_1(\theta, q_2^{\rm full})\right)\right]\\
        &- \Phi\!\left[\min\!\left(B_1(\theta),R_1(\theta, q_1^{\rm crit})\right)\right],
    \end{aligned}
\end{equation}
where $q_1^{\rm crit}$ satisfies equation \eqref{eq:qcrit}.

\section{Expected upper limits}
\label{app:expected}

In this section, we provide explicit expressions for the expected upper limits $\mathbb{E}[\TUL]$, for the frameworks considered in this work and $N_s\le 2$.

In the geometric construction $\TUL^{g_{N_s}}$ does not depend on the magnitude of the observed data. Hence, the value provided by \eqref{eq:def:geomUL} coincides with the expected benchmark.

For the likelihood framework with $N_s = 1$, we derived $\TUL^{\ell1 }=s_1+k_{\alpha}$, where we defined $k_{\alpha}=\Phi^{-1}(1-\alpha)$. Replacing $s_1$ with its expected value $\theta$ immediately gives $\mathbb{E}[\TUL^{\ell1}]$. In the constrained case:
\begin{equation}\label{eq:integ:ful1p}
    \mathbb{E}[\TUL^{\ell 1+}] = \int_{-k_\alpha}^{\infty}
    (s_1 + k_\alpha)\,\phi(s_1-\theta)\,\mathrm{d}s_1,
\end{equation}
where $\phi(s_1-\theta)$ is the distribution of $s_1$ and $\phi$ is the standard normal CDF. Integration of \eqref{eq:integ:ful1p} gives
\begin{equation}\label{eq:Aul:Exp1Const}
    \mathbb{E}[\TUL^{\ell1+}] =
    (\theta + k_\alpha) \, \Phi(\theta + k_\alpha)
    + \phi(\theta + k_\alpha).
\end{equation}

Two branches contribute to $\mathbb{E}[\TUL^{\ell 2}] = E_1 + E_2$.
Similarly to \eqref{eq:integ:ful1p}, $E_1$ comes from the one-scan outcome:
\begin{equation}\label{eq:def:F1UL}
    E_1 = \int_{-\infty}^\beta
    \TUL(s_1)\,\phi(s_1 - \theta)\,\mathrm{d}s_1,
\end{equation}
where $\TUL(s_1)$ solves $p_\vartheta^{(2)}=\alpha$, and has to be evaluated numerically from equation \eqref{eq:def:p2}. In the constrained case, $\TUL^{+}(s_1)$ is the upper endpoint of the region $p_{\vartheta^+_{m=1}}^{(2)}\ge\alpha$:
\begin{equation}\label{eq:ineq:th1+}
    \TUL^{+}(s_1)=\sup\left\{\vartheta^+_{m=1}\ge0:p_{\vartheta^+_{m=1}}^{(2)}\ge\alpha\right\}.
\end{equation}
Inequality \eqref{eq:ineq:th1+} handles the case of strongly negative realization, for which there might not be a positive root.

For the contribution from the second branch, we have
\begin{equation}\label{eq:def:F2}
    E_2 =\int_{-\infty}^{\infty}\TUL(s_2)\,
    f(s_2)\,\Prob(S_1>\beta\mid S_2 = s_2)\,\mathrm{d}s_2,
\end{equation}
where $\TUL(s_2)$ differs from $\TUL(s_1)$ in that the unconstrained MLE is now $\hat{\vartheta}=\rho_{12} s_2\equiv\rho s_2$. The value $\vartheta(s_2)$ (and its constrained counterpart) are once again obtained from numerically solving the condition over $p_\vartheta^{(2)}$.

The distribution of $s_2$, $f(s_2)$, comes directly from \eqref{eq:def:S}:
\begin{equation}
    f(s_2)=\mathcal{N}(\theta\sqrt{1+\omega},1)=\phi(s_2-\theta/\rho).
\end{equation}
In order to compute $\Prob(S_1>\beta\mid S_2 = s_2)$ we need the distribution of $S_1\mid S_2 = s_2$. We apply the standard formulae for a partitioned multivariate normal distribution to $\bm{S}=(S_1, S_2)$---see e.g.\ \cite[Sec.\ 2.5]{Anderson2003}. This gives:
\begin{align}
    S_1\,|\,S_2 &=
    s_{2}\sim\mathcal{N}(\mu_{s_1}+\rho(s_{2}-\mu_{s_2}), 1-\rho^2)\\
    &=
    \mathcal{N}(\rho s_{2}, 1-\rho^2),
\end{align}
and therefore:
\begin{align}
    \Prob(S_1>\beta\mid S_2 = s_2)
    &=1-\Phi\!\left(\frac{\beta-\rho s_2}{\sqrt{1-\rho^2}}\right).
\end{align}

The same line of reasoning applies to $N_s>2$ and to $\mathbb{E}[\TUL^{\rm mix}]$. In the latter case, contribution $E_1$ \eqref{eq:def:F1UL} from the one-scan branch would just require $\TUL(s_1)$ to be substituted with $\TUL^{\omega\to\infty}(s_1)$.

Concerning the BPM framework, we combine equation (9) of Ref.\ \cite{Palken:2020wgs} with the definition $\lambda = \exp(-\kappa^2/2)$. This allows to express the exclusion condition $\Lambda(\vartheta)\leq\lambda$ on the standardized variable $Z_m$ as $Z_m\leq R_m^\Lambda$, where
\begin{equation}\label{eq:def:RL}
    R_1^\Lambda =
    \frac{\vartheta-2\theta}{2} -\frac{\kappa^2}{2\vartheta}
    \quad\text{and}\quad
    R_2^\Lambda =
    \frac{\vartheta-2\theta}{2\rho}
    -\frac{\rho\kappa^2}{2\vartheta}.
\end{equation}

For BPM-fix, exclusion occurs either after stopping at the first scan, $\{Z_1(\theta)<B_1(\theta),\;Z_1(\theta)\leq R_1^\Lambda\}$, or after a triggered rescan, $\{Z_1(\theta)\geq B_1(\theta),\;Z_2(\theta)\leq R_2^\Lambda\}$. For BPM-any, instead, a tested value is excluded as soon as the BPM cut is crossed, that is,\ $\{Z_1(\theta)\leq R_1^\Lambda\}$ or $\{Z_1(\theta)\geq B_1(\theta),\;Z_2(\theta)\leq R_2^\Lambda\}$. Therefore, the probabilities associated to those events are:
\begin{align}
 E_\theta^{\rm fix}(\vartheta)
 &=\Phi\!\left[\min(B_1,R_1^\Lambda)\right]
   +\Phi(R_2^\Lambda)
   -\Phi_2(B_1,R_2^\Lambda;\rho),\nonumber\\
 E_\theta^{\rm any}(\vartheta)
 &=\Phi(R_1^\Lambda)+\Phi(R_2^\Lambda)
   -\Phi_2\!\left[\max(R_1^\Lambda,B_1),
                  R_2^\Lambda;\rho\right].\nonumber
\end{align}

Finally, since $E_\theta^k(\vartheta)=P_\theta(\TUL^k\le\vartheta)$, its complement is the survival function of the upper limit. Hence:
\begin{equation}
    \mathbb E_\theta[\TUL^k]=
    \int_0^\infty \left[1-E_\theta^k(\vartheta)\right]\,\mathrm{d}\vartheta,
\end{equation}
with $k=$ fix, any.
The two implementations differ only when the first scan already satisfies $\Lambda(\vartheta)\leq\lambda$, triggers a rescan and then the second scan reverses that exclusion. The probability of this happening is
\begin{equation}\label{eq:com:Pth}
    \Prob_\theta\!\left(\beta < S_1\le
    \frac{\vartheta}{2} -\frac{\kappa^2}{2\vartheta},\,
    S_2 > \frac{\vartheta}{2\rho}
    -\frac{\rho\kappa^2}{2\vartheta}
    \right),
\end{equation}
and the two implementations coincides for $2\beta> \vartheta -\kappa^2/\vartheta$.

\printbibliography

@book{Anderson2003,
  author    = {Anderson, Theodore W.},
  title     = {An Introduction to Multivariate Statistical Analysis},
  edition   = {3},
  series    = {Wiley Series in Probability and Statistics},
  publisher = {John Wiley \& Sons},
  address   = {Hoboken, NJ},
  year      = {2003},
  isbn      = {978-0-471-36091-9}, 
  doi       = {10.1002/0471722065}
}

@article{Grunwald:2024,
    author = {Grünwald, Peter and de Heide, Rianne and Koolen, Wouter},
    title = {Safe testing},
    journal = {Journal of the Royal Statistical Society Series B: Statistical Methodology},
    volume = {86},
    number = {5},
    pages = {1091-1128},
    year = {2024},
    month = {11},
    issn = {1369-7412},
    doi = {10.1093/jrsssb/qkae011},
    url = {https://doi.org/10.1093/jrsssb/qkae011},
    eprint = {https://academic.oup.com/jrsssb/article-pdf/86/5/1091/60648648/qkae011.pdf},
}

@book{Ville:1939,
  author    = {Ville, Jean},
  title     = {\'{E}tude critique de la notion de collectif},
  year      = {1939},
  publisher = {Gauthier-Villars},
  series    = {Monographies des probabilit\'{e}s},
  volume    = {3},
  address   = {Paris},
  url       = {http://eudml.org/doc/192893}
}

@article{RamdasWang:2025,
    author = {Ramdas, Aaditya and Wang, Ruodu},
    title = {Hypothesis Testing with E-values},
    journal = {Foundations and Trends in Statistics},
    volume = {1},
    number = {1-2},
    pages = {1-390},
    year = {2025},
    month = {07},
    issn = {2978-4212},
    doi = {10.1561/3600000002},
    url = {https://doi.org/10.1561/3600000002},
    eprint = {https://www.emerald.com/ftstat/article-pdf/1/1-2/1/11146063/3600000002en.pdf},
}

@phdthesis{Brubaker:2017ohw,
    author = "Brubaker, B. M.",
    title = "{First results from the HAYSTAC axion search}",
    eprint = "1801.00835",
    archivePrefix = "arXiv",
    school = "Yale U.",
    year = "2017"
}

@article{Brubaker:2017rna,
    author = "Brubaker, B. M. and Zhong, L. and Lamoreaux, S. K. and Lehnert, K. W. and van Bibber, K. A.",
    title = "{HAYSTAC axion search analysis procedure}",
    eprint = "1706.08388",
    archivePrefix = "arXiv",
    primaryClass = "astro-ph.IM",
    doi = "10.1103/PhysRevD.96.123008",
    journal = "Phys. Rev. D",
    volume = "96",
    number = "12",
    pages = "123008",
    year = "2017"
}

@article{Foster:2017hbq,
    author = "Foster, Joshua W. and Rodd, Nicholas L. and Safdi, Benjamin R.",
    title = "{Revealing the Dark Matter Halo with Axion Direct Detection}",
    eprint = "1711.10489",
    archivePrefix = "arXiv",
    primaryClass = "astro-ph.CO",
    reportNumber = "MIT-CTP-4964, LCTP-17-08, MIT-CTP 4964",
    doi = "10.1103/PhysRevD.97.123006",
    journal = "Phys. Rev. D",
    volume = "97",
    number = "12",
    pages = "123006",
    year = "2018"
}

@article{Palken:2020wgs,
    author = "Palken, D. A. and others",
    title = "{Improved analysis framework for axion dark matter searches}",
    eprint = "2003.08510",
    archivePrefix = "arXiv",
    primaryClass = "astro-ph.IM",
    doi = "10.1103/PhysRevD.101.123011",
    journal = "Phys. Rev. D",
    volume = "101",
    number = "12",
    pages = "123011",
    year = "2020"
}

@article{GalloRosso:2022mhx,
    author = "Gallo Rosso, Andrea and Algeri, Sara and Conrad, Jan",
    title = "{Sequential hypothesis testing for axion haloscopes}",
    eprint = "2210.16095",
    archivePrefix = "arXiv",
    primaryClass = "physics.data-an",
    doi = "10.1103/PhysRevD.108.023003",
    journal = "Phys. Rev. D",
    volume = "108",
    number = "2",
    pages = "023003",
    year = "2023"
}

@article{NeymanPearson1933,
  title={On the problems of the most efficient tests of statistical hypotheses},
  author={Neyman, J. and Pearson, E. S.},
  journal={Philosophical Transactions of the Royal Society of London},
  volume={231A},
  pages={289--338},
  year={1933},
  doi={10.1098/rsta.1933.0009},
  url={https://doi.org/10.1098/rsta.1933.0009}
}

@book{Genz:2009,
  title     = {Computation of Multivariate Normal and $t$ Probabilities},
  author    = {Genz, Alan and Bretz, Frank},
  series    = {Lecture Notes in Statistics},
  volume    = {195},
  year      = {2009},
  publisher = {Springer-Verlag},
  address   = {Berlin, Heidelberg},
  doi       = {10.1007/978-3-642-01689-9},
  isbn      = {978-3-642-01688-2}
}

@book{Feller:1968,
  title     = {An Introduction to Probability Theory and Its Applications},
  author    = {Feller, William},
  volume    = {1},
  edition   = {3rd},
  year      = {1968},
  publisher = {John Wiley \& Sons},
  address   = {New York},
  isbn      = {978-0-471-25708-0}
}

@article{Ahn:2020nyv,
    author = "Ahn, S. and Lee, S. and Choi, J. and Ko, B. R. and Semertzidis, Y. K.",
    title = "{Improved axion haloscope search analysis}",
    eprint = "2004.08011",
    archivePrefix = "arXiv",
    primaryClass = "astro-ph.CO",
    doi = "10.1007/JHEP04(2021)297",
    journal = "JHEP",
    volume = "04",
    pages = "297",
    year = "2021"
}

@article{Knirck:2018knd,
    author = "Knirck, Stefan and Millar, Alexander J. and O'Hare, Ciaran A. J. and Redondo, Javier and Steffen, Frank D.",
    title = "{Directional axion detection}",
    eprint = "1806.05927",
    archivePrefix = "arXiv",
    primaryClass = "astro-ph.CO",
    doi = "10.1088/1475-7516/2018/11/051",
    journal = "JCAP",
    volume = "11",
    pages = "051",
    year = "2018"
}

@article{Irastorza:2018dyq,
    author = "Irastorza, Igor G. and Redondo, Javier",
    title = "{New experimental approaches in the search for axion-like particles}",
    doi = "10.1016/j.ppnp.2018.05.003",
    journal = "Prog.\ Part.\ Nucl.\ Phys.",
    volume = "102",
    pages = "89--159",
    year = "2018",
    eprint = "1801.08127",
    archivePrefix = "arXiv",
    primaryClass = "hep-ph"
}

@article{Billard:2021uyg,
    author = "Billard, J. and others",
    title = "{Direct detection of dark matter -- APPEC committee report}",
    doi = "10.1088/1361-6471/ac5754",
    journal = "J.\ Phys.\ G",
    volume = "49",
    number = "4",
    pages = "040501",
    year = "2022",
    eprint = "2104.07634",
    archivePrefix = "arXiv",
    primaryClass = "hep-ex"
}

@article{Irastorza:2021tdu,
    author = "Irastorza, Igor G.",
    title = "{Laboratory searches for galactic axions}",
    doi = "10.1088/1361-6633/ac0d1c",
    journal = "Rept.\ Prog.\ Phys.",
    volume = "84",
    number = "10",
    pages = "106901",
    year = "2021",
    eprint = "2103.12400",
    archivePrefix = "arXiv",
    primaryClass = "hep-ph"
}

@article{Semertzidis:2021rxs,
    author = "Semertzidis, Yannis K. and Youn, Seongtae",
    title = "{Axion dark matter: How to see it?}",
    doi = "10.21468/SciPostPhysProc.2.021",
    journal = "SciPost Phys.\ Proc.",
    volume = "2",
    pages = "021",
    year = "2019",
    eprint = "1904.04168",
    archivePrefix = "arXiv",
    primaryClass = "hep-ph"
}

@article{Berlin:2024pzi,
    author = "Berlin, Asher and Kahn, Yonatan",
    title = "{New Technologies for Axion and Dark Photon Searches}",
    eprint = "2412.08704",
    archivePrefix = "arXiv",
    primaryClass = "hep-ph",
    reportNumber = "FERMILAB-PUB-24-0933-T",
    doi = "10.1146/annurev-nucl-121423-101015",
    journal = "Ann. Rev. Nucl. Part. Sci.",
    volume = "75",
    number = "1",
    pages = "83--108",
    year = "2025"
}

@proceedings{Adams:2855525,
    editor = "Adams, Douglas and others",
    title = "{Axion Dark Matter}",
    reportNumber = "CERN-2021-001",
    doi = "10.17181/CERN.97QY-CKH9",
    publisher = "CERN",
    address = "Geneva",
    year = "2021",
    series = "CERN Yellow Reports: School Proceedings",
    volume = "1"
}

@book{AxioBook2022,
    editor = "Marsh, David J. E. and Irastorza, Igor G. and Redondo, Javier",
    title = "{The Physics of Axions}",
    doi = "10.1007/978-3-030-95852-7",
    publisher = "Springer",
    address = "Cham",
    year = "2022"
}

@article{Preskill:1982cy,
    author = "Preskill, J. and Wise, M. B. and Wilczek, F.",
    editor = "Srednicki, M. A.",
    title = "{Cosmology of the Invisible Axion}",
    reportNumber = "HUTP-82-A048, NSF-ITP-82-103",
    doi = "10.1016/0370-2693(83)90637-8",
    journal = "Phys. Lett. B",
    volume = "120",
    pages = "127--132",
    year = "1983"
}

@article{Abbott:1982af,
    author = "Abbott, L. F. and Sikivie, P.",
    editor = "Srednicki, M. A.",
    title = "{A Cosmological Bound on the Invisible Axion}",
    reportNumber = "PRINT-82-0695 (BRANDEIS)",
    doi = "10.1016/0370-2693(83)90638-X",
    journal = "Phys. Lett. B",
    volume = "120",
    pages = "133--136",
    year = "1983"
}

@article{Dine:1982ah,
    author = "Dine, M. and Fischler, W.",
    editor = "Srednicki, M. A.",
    title = "{The Not So Harmless Axion}",
    reportNumber = "UPR-0201T",
    doi = "10.1016/0370-2693(83)90639-1",
    journal = "Phys. Lett. B",
    volume = "120",
    pages = "137--141",
    year = "1983"
}

@misc{Green:2022hhj,
    author = "Green, Daniel and others",
    title = "{Snowmass Theory Frontier: Astrophysics and Cosmology}",
    eprint = "2209.06854",
    archivePrefix = "arXiv",
    primaryClass = "hep-ph",
    reportNumber = "FERMILAB-PUB-22-721-T",
    month = sep,
    year = "2022"
}

@article{Arbey:2021gdg,
    author = "Arbey, A. and Mahmoudi, F.",
    title = "{Dark matter and the early Universe: a review}",
    eprint = "2104.11488",
    archivePrefix = "arXiv",
    primaryClass = "hep-ph",
    reportNumber = "CERN-TH-2021-066",
    doi = "10.1016/j.ppnp.2021.103865",
    journal = "Prog. Part. Nucl. Phys.",
    volume = "119",
    pages = "103865",
    year = "2021"
}

@article{Profumo:2019ujg,
    author = "Profumo, S. and Giani, L. and Piattella, O. F.",
    title = "{An Introduction to Particle Dark Matter}",
    eprint = "1910.05610",
    archivePrefix = "arXiv",
    primaryClass = "hep-ph",
    doi = "10.3390/universe5100213",
    journal = "Universe",
    volume = "5",
    number = "10",
    pages = "213",
    year = "2019"
}

@article{Feng:2010gw,
    author = "Feng, J. L.",
    title = "{Dark Matter Candidates from Particle Physics and Methods of Detection}",
    eprint = "1003.0904",
    archivePrefix = "arXiv",
    primaryClass = "astro-ph.CO",
    reportNumber = "UCI-TR-2009-13",
    doi = "10.1146/annurev-astro-082708-101659",
    journal = "Ann. Rev. Astron. Astrophys.",
    volume = "48",
    pages = "495--545",
    year = "2010"
}

@article{Jaeckel:2010ni,
    author = "Jaeckel, J. and Ringwald, A.",
    title = "{The Low-Energy Frontier of Particle Physics}",
    eprint = "1002.0329",
    archivePrefix = "arXiv",
    primaryClass = "hep-ph",
    reportNumber = "CPT-10-18, DESY-10-016, IPPP-10-09",
    doi = "10.1146/annurev.nucl.012809.104433",
    journal = "Ann. Rev. Nucl. Part. Sci.",
    volume = "60",
    pages = "405--437",
    year = "2010"
}

@article{Bertone:2004pz,
    author = "Bertone, G. and Hooper, D. and Silk, J.",
    title = "{Particle dark matter: Evidence, candidates and constraints}",
    eprint = "hep-ph/0404175",
    archivePrefix = "arXiv",
    reportNumber = "FERMILAB-PUB-04-047-A",
    doi = "10.1016/j.physrep.2004.08.031",
    journal = "Phys. Rept.",
    volume = "405",
    pages = "279--390",
    year = "2005"
}

@article{Bergstrom:2000pn,
    author = {Bergstr\"om, L.},
    title = "{Nonbaryonic dark matter: Observational evidence and detection methods}",
    eprint = "hep-ph/0002126",
    archivePrefix = "arXiv",
    doi = "10.1088/0034-4885/63/5/2r3",
    journal = "Rept. Prog. Phys.",
    volume = "63",
    pages = "793",
    year = "2000"
}

@article{Zhitnitsky:1980tq,
  author = {Zhitnitsky, A. R.},
  title = {{On Possible Suppression of the Axion Hadron Interactions. (In Russian)}},
  journal = {Sov.\ J.\ Nucl.\ Phys.},
  volume = {31},
  pages = {260},
  year = {1980},
}

@article{Dine:1981rt,
  author = {Dine, Michael and Fischler, Willy and Srednicki, Mark},
  title = {{A Simple Solution to the Strong CP Problem with a Harmless Axion}},
  doi = {10.1016/0370-2693(81)90590-6},
  journal = {Phys.\ Lett.\ B},
  volume = {104},
  pages = {199--202},
  year = {1981},
}

@article{Shifman:1979if,
  author = {Shifman, M. A. and Vainshtein, A. I. and Zakharov, V. I.},
  title = {{Can Confinement Ensure Natural CP Invariance of Strong Interactions?}},
  doi = {10.1016/0370-2693(80)90209-6},
  journal = {Phys.\ Lett.\ B},
  volume = {78},
  pages = {443--446},
  year = {1978},
}

@article{Kim:1979if,
  author = {Kim, Jihn E.},
  title = {{Weak Interaction Singlet and Strong CP Invariance}},
  doi = {10.1103/PhysRevLett.43.103},
  journal = {Phys.\ Rev.\ Lett.},
  volume = {43},
  pages = {103},
  year = {1979},
}

@article{Wilczek:1977pj,
  author = {Wilczek, Frank},
  title = {{Problem of Strong P and T Invariance in the Presence of Instantons}},
  doi = {10.1103/PhysRevLett.40.279},
  journal = {Phys.\ Rev.\ Lett.},
  volume = {40},
  pages = {279--282},
  year = {1978},
}

@article{Weinberg:1977ma,
  author = {Weinberg, Steven},
  title = {{A New Light Boson?}},
  doi = {10.1103/PhysRevLett.40.223},
  journal = {Phys.\ Rev.\ Lett.},
  volume = {40},
  pages = {223--226},
  year = {1978},
}

@article{Peccei:1977ur,
  author = {Peccei, R. D. and Quinn, Helen R.},
  title = {{Constraints Imposed by CP Conservation in the Presence of Instantons}},
  reportnumber = {ITP-572-STANFORD},
  doi = {10.1103/PhysRevD.16.1791},
  journal = {Phys.\ Rev.\ D},
  volume = {16},
  pages = {1791--1797},
  year = {1977},
}

@article{Peccei:1977hh,
  author = {Peccei, R. D. and Quinn, Helen R.},
  title = {{CP Conservation in the Presence of Instantons}},
  reportnumber = {ITP-568-STANFORD},
  doi = {10.1103/PhysRevLett.38.1440},
  journal = {Phys.\ Rev.\ Lett.},
  volume = {38},
  pages = {1440--1443},
  year = {1977},
}

@article{Sikivie:1983ip,
    author = "Sikivie, P.",
    editor = "Srednicki, M. A.",
    title = "{Experimental Tests of the Invisible Axion}",
    reportNumber = "PRINT-83-0597 (FLORIDA), UF-TP-83-13",
    doi = "10.1103/PhysRevLett.51.1415",
    journal = "Phys. Rev. Lett.",
    volume = "51",
    pages = "1415--1417",
    year = "1983",
    note = "[Erratum: Phys.Rev.Lett. 52, 695 (1984)]"
}

@article{Sikivie:1985yu,
    author = "Sikivie, Pierre",
    title = "{Detection Rates for 'Invisible' Axion Searches}",
    reportNumber = "UFTP-85-5-REV, UFTP-85-5",
    doi = "10.1103/PhysRevD.36.974",
    journal = "Phys. Rev. D",
    volume = "32",
    pages = "2988",
    year = "1985",
    note = "[Erratum: Phys.Rev.D 36, 974 (1987)]"
}

@misc{ALPHA:2026vwc,
    author = "Bai, Xiran and others",
    collaboration = "ALPHA",
    title = "{Design of ALPHA Phase I: A Plasma Haloscope for 10--20 GHz Post-Inflation Axions}",
    eprint = "2608.10292",
    archivePrefix = "arXiv",
    primaryClass = "hep-ex",
    month = "8",
    year = "2026"
}

@book{jeffreys1998,
  author       = {Jeffreys, Harold},
  title        = {The Theory of Probability},
  edition      = {3},
  year         = {1998},
  origyear     = {1961},
  publisher    = {Oxford University Press},
  location     = {Oxford, England},
  pagetotal    = {432},
  isbn         = {978-0-19-158967-6}
}

@article{kass1995,
  author  = {Kass, Robert E. and Raftery, Adrian E.},
  title   = {Bayes Factors},
  journal = {Journal of the American Statistical Association},
  year    = {1995},
  volume  = {90},
  number  = {430},
  pages   = {773--795},
  doi     = {10.1080/01621459.1995.10476572}
}

\end{document}